\def\Fint{\rlap{$\Biggl\rfloor$}\Biggl\lceil}
\def\sla{\raise.15ex\hbox{$/$}\kern-.57em}
\def\slas#1{\rlap{$#1$}\thinspace /}

\documentstyle[epsfig,12pt]{article}

\begin{document}
\begin{titlepage}
\begin{flushright}
hep-th/0110180 \\ Brown-HET-1289 \\ UFIFT-HEP-01-14
\end{flushright}
\begin{center}
\textbf{The One Loop Effective Action of QED \\ 
for a General Class of Electric Fields}
\end{center}
\begin{center}
H. M. Fried$^{\dagger}$
\end{center}
\begin{center}
\textit{Department of Physics \\ Brown University \\ Providence, RI 02912 \\ 
UNITED STATES}
\end{center}
\begin{center}
R. P. Woodard$^{\ddagger}$
\end{center}
\begin{center}
\textit{Department of Physics \\ University of Florida \\ 
Gainesville, FL 32611 \\ UNITED STATES}
\end{center}
\begin{center}
ABSTRACT
\end{center}
\hspace*{.5cm} We compute the effective action of QED at one loop order for 
an electric field which points in the $\widehat{z}$ direction and depends 
arbitrarily upon the light cone time coordinate, $x^+ = (x^0 + x^3)/\sqrt{2}$.
This calculation generalizes Schwinger's formula for the vacuum persistence 
probability in the presence of a constant electric field.
\begin{flushleft}
\end{flushleft}
\begin{flushleft}
$^{\dagger}$ e-mail: fried@het.brown.edu \\
$^{\ddagger}$ e-mail: woodard@phys.ufl.edu
\end{flushleft}
\end{titlepage}

Schwinger's 1951 calculation \cite{Schwinger} of the vacuum persistence
probability $P_0$ in the presence of a constant electric field has stood 
the test of time. For half a century it has been the exact calculation to 
which other, subsequent models of complete solubility have been compared 
\cite{HoNa}. Crude model estimates of pair production in the overlap region 
of two high-intensity lasers have recently been given \cite{FGMA}. As in 
all previous expressions for $P_0$ these estimates contain an essential 
singularity in the coupling constant, suggesting that such production 
processes may be designated as ``intrinsically non-perturbative''. We 
prefer this designation to the ubiquitous phrase, ``non-Borel summable''. 

A significant and exactly soluble generalization of Schwinger's result has
recently appeared \cite{TTW1,TTW2}, in which the electric field $\vec{E} =
\widehat{z} E(x^+)$ is parallel to the $z$ axis and can depend arbitrarily 
upon the light cone time parameter, $x^+\equiv (x^0 + x^3)/\sqrt{2}$.\footnote{
Summaries of the material in ref. \cite{FGMA} and \cite{TTW1,TTW2} were 
recently presented at the Sixth Workshop on Non-Perturbative QCD, at the 
American University of Paris, France, June 5-9, 2001.} In that work the 
Heisenberg field equations were solved for the Fermion operators in the 
presence of the electric background; then these solutions were used to 
compute $P_0$ and the one loop expectation values of the various bilinears. 
In this note we describe an alternate, functional treatment in which the 
one loop effective action is computed for this same background field.

The feature of the background which makes our treatment possible is the fact
that it depends only upon either $x^+$ or $x^- \equiv (x^0 - x^3)/\sqrt{2}$; it
does not matter which. For definiteness we shall assume dependence upon $x^+$,
corresponding to the following vector potential,
\begin{equation}
A_-(x^+) = \int_0^{x^+} du E(u) \; .
\end{equation}
We work in the gauge $A_+ = 0$ and we assume that the transverse components
of the vector potential vanish, $A_{\bot} = 0$.

The vacuum persistence probability is the norm-squared of the vacuum-to-vacuum
amplitude in the presence of background $A_{\mu}$,
\begin{equation}
P_0[A] = \left\Vert e^{i \Gamma[A]} \right\Vert^2 \; .
\end{equation}
The fermionic contribution to the vacuum-to-vacuum amplitude can be written 
as the ratio of functional determinants,
\begin{equation}
e^{L[A+a]} = {{\rm det}(i \sla{\partial} - e \slas{A}  - e \sla{a} - m) \over 
{\rm det}(i \sla{\partial} - m)} \; ,
\end{equation}
where $a_{\mu}$ is the quantum gauge field. Functionally integrating over 
$a_{\mu}$ gives the full effective action. This can be written using the
functional notation of ref. \cite{Fried1},
\begin{equation}
e^{i\Gamma[A]} = \left. e^{{\cal D}_A} \cdot e^{L[A+a]} \right\vert_{a_{\mu}
=0} \; .
\end{equation}
Here the exponent of the functional linkage operator is,
\begin{equation}
{\cal D}_A \equiv -\frac{i}2 \int d^4x \int d^4y {\delta \over \delta 
a_{\mu}(x)} D_{\mu\nu}(x-y) {\delta \over \delta a_{\nu}(y)} \; .
\end{equation}
and $D_{\mu\nu}(x-y)$ is the free photon propagator. The linkage gives higher
loop corrections. Only the fermionic determinant is needed to the one loop 
order that we are working.

It will be convenient to employ Fradkin's representation \cite{Fradkin} for 
the logarithm of the fermionic determinant, specialized to our 
background,
\begin{eqnarray}
\lefteqn{L[A] = -\frac12 \int d^4x \int_0^{\infty} {ds \over s} e^{-i s m^2}
\int {d^4p \over (2\pi)^4} \exp\left[i \int_0^s ds' {\delta \over \delta
v_{\mu}(s')} {\delta \over \delta v^{\mu}(s')} \right]} \nonumber \\
& & \times \exp\left[ i \int_0^s ds' v_{\mu}(s') p^{\mu} \right] \left\{ 
\exp\left[ i e \int_0^s ds' v_+(s') A_-\left(x^+ + \int_0^{s'} ds'' v_-(s'') 
\right) \right] \right. \nonumber \\
& & \qquad \times \left. {\rm Tr}\left( \exp\left[i e \int_0^s ds' \sigma^{03} 
E\left( x^+ + \int_0^{s'} ds'' v_-(s'') \right)\right]\right) - 1 \right\}_{
v_{\mu} \rightarrow 0} \; .
\end{eqnarray}
(For a derivation of Fradkin's representation with applications, see ref. 
\cite{Fried2}.) The first step in evaluating this expression is the 
introduction of a functional integral representation of unity,
\begin{equation}
1 = \Fint d[u] \delta\left[u(s') + \int_0^{s'} ds'' v_-(s'')\right] \; .
\end{equation}
We use this to extract $v_-$ dependence from the background fields,
\begin{eqnarray}
\lefteqn{L[A] = -2 \int d^4x \int_0^{\infty} {ds \over s} e^{-i s m^2}
\int {d^4p \over (2\pi)^4} \exp\left[i \int_0^s ds' {\delta \over \delta
v_{\mu}(s')} {\delta \over \delta v^{\mu}(s')} \right]} \nonumber \\
& & \times \exp\left[ i \int_0^s ds' v_{\mu}(s') p^{\mu} \right] \Fint d[u] 
\delta\left[u(s') + \int_0^{s'} ds'' v_-(s'')\right] \nonumber \\
& & \hspace{1.5cm} \times \left\{ \exp\left[ i e \int_0^s ds' v_+(s') 
A_-\left(x^+ - u(s') \right) \right] \right. \nonumber \\
& & \hspace{3cm} \left. \cosh\left[ \int_0^s ds' e E\left(x^+ - u(s')\right) 
\right] - 1 \right\}_{v_{\mu} \rightarrow 0} \; .
\end{eqnarray}

The next step is substituting a Fourier representation for the delta 
functional,
\begin{eqnarray}
\lefteqn{\delta\left[u(s') + \int_0^{s'} ds'' v_-(s'')\right]} \nonumber \\
& & = \Fint d[\Omega] \exp\left[i \int_0^s ds' \Omega(s')
\left(u(s') + \int_0^s ds'' v_-(s'') \right) \right] \; , \\
& & = \Fint d[\Omega] \exp\left[i \int_0^s ds' \Omega(s') u(s')\right]
\exp\left[i \int_0^s ds' v_-(s') \int_{s'}^s ds'' \Omega(s'') \right] \; .
\qquad
\end{eqnarray}
We can now explicitly evaluate the functional linkage carried by $v_{\mu}(s')$,
\begin{eqnarray}
\lefteqn{\exp\left[i \int_0^s ds' {\delta \over \delta v_{\mu}(s')} 
{\delta \over \delta v^{\mu}(s')}\right] \exp\left[i \int_0^s ds' \left(p^{\mu} 
v_{\mu}(s') {\mbox{} \over \mbox{}} \right. \right.} \nonumber \\
& & \hspace{2.5cm} \left. \left. \left. + v_+(s') e A_-(x^+ - u(s')) + v_-(s')
\int_{s'}^s ds'' \Omega(s'') \right) \right] \right\vert_{v_{\mu} \rightarrow 
0} = \nonumber \\
& & e^{-i s p_{\bot}^2} \exp\left[2i \int_0^s ds' \left\{p^+ + e A_-(x^+ - 
u(s'))\right\} \left\{p^- + \int_{s'}^s ds'' \Omega(s'')\right\} \right] , \\
& & = e^{-i s p^2} \exp\left[2i p^- \int_0^s ds' e A_-(x^+ - u(s')) + 2i p^+
\int_0^s ds' s' \Omega(s') \right. \nonumber \\
& & \hspace{4cm} \left. + 2i \int_0^s ds' e A_-(x^+ - u(s')) \int_{s'}^s ds'' 
\Omega(s'')\right] \; .
\end{eqnarray}
The resulting expression for the logarithm of the fermionic determinant is,
\begin{eqnarray}
\lefteqn{L[A] = -2 \int d^4x \int_0^{\infty} {ds \over s} e^{-i s m^2} \int 
{d^4p \over (2\pi)^4} e^{-i s p^2 } } \nonumber \\
& & \times \Fint d[u] \Fint d[\Omega] \exp\left[i \int_0^s ds' \Omega(s') 
\left( u(s') + 2 s' p^+\right)\right] \nonumber \\
& & \hspace{1.5cm} \times \left\{ \exp\left[ 2 i \int_0^s ds' e A_-\left(x^+ 
- u(s') \right) \left(p^+ + \int_0^{s'} ds'' \Omega(s'') \right)\right] \right.
\nonumber \\
& & \hspace{3cm} \left. \times \cosh\left[ \int_0^s ds' e E\left(x^+ - 
u(s')\right) \right] - 1 \right\} \; .
\end{eqnarray}

Performing the functional integration over $\Omega$ gives functional delta 
functions. The $e=0$ one of these can be done trivially,
\begin{eqnarray}
\lefteqn{L[A] = -2 \int d^4x \int_0^{\infty} {ds \over s} e^{-i s m^2} \int 
{d^4p \over (2\pi)^4} e^{-i s p^2 } } \nonumber \\
& & \times \left\{ \Fint d[u] \delta\left[ u(s') + 2 s' p^+ + 2 \int_0^{s'}
ds'' e A_-\left(x^+ - u(s'') \right) \right] \right. \\
& & \left. \times \exp\left[i 2 p^- \int_0^s ds' e A_-\left(x^+ - u(s')\right) 
\right] \cosh\left[ \int_0^s ds' e E\left(x^+ - u(s')\right) \right] - 
1 \right\} \nonumber
\end{eqnarray}
The $e$-dependent delta functional is not so simple to evaluate. It evidently 
requires solution of the integral equation,
\begin{equation}
u(s') + 2 s' p^+ + 2 \int_0^{s'} ds'' e A_-\left(x^+ - u(s'')\right) = 0 \; .
\end{equation}

We make further progress by performing the momentum integrations. The 
transverse ones give a simple factor,
\begin{equation}
\int {d^2 p_{\bot} \over (2\pi)^2} e^{-i s p_{\bot} \cdot p_{\bot}}
= {-i \over 4 \pi s} \; .
\end{equation}
It is the one over $p^-$ which provides the essential simplification,
\begin{eqnarray}
\lefteqn{\int_{-\infty}^{\infty} {dp^- \over 2\pi} e^{2i s p^+ p^-} 
\exp\left[i 2 p^- \int_0^s ds' e A_-\left(x^+ - u(s')\right)\right] } 
\nonumber \\
& & \hspace{3.5cm} = \frac1{2s} \delta\left( p^+ + \frac1{s} \int_0^s ds' 
eA_-\left(x^+ - u(s')\right)\right) \; .
\end{eqnarray}
Evaluating the trivial $p^+$ integral results in the following form for the
logarithm of the fer\-mi\-on\-ic determinant,
\begin{eqnarray}
\lefteqn{L[A] = {i \over 8 \pi^2} \int d^4x \int_0^{\infty} {ds \over s^3} 
e^{-i s m^2} \left\{ \Fint d[u] \delta\left[ u(s') - 2 \frac{s'}{s} \int_0^s 
ds'' e A_-\left(x^+ - u(s'')\right) \right. \right.} \nonumber \\
& & \left. \left. + 2 \int_0^{s'} ds'' e A_-\left(x^+ - u(s'') \right) \right] 
\cosh\left[ \int_0^s ds' e E\left(x^+ - u(s')\right) \right] - 1 \right\} \; .
\qquad \label{neweq}
\end{eqnarray}

It might seem that fixing $p^+$ has actually made the integral equation more
difficult to solve,
\begin{equation}
u(s') - 2 \frac{s'}{s} \int_0^s ds'' e A_-\left(x^+ - u(s'')\right) 
+ 2 \int_0^{s'} ds'' e A_-\left(x^+ - u(s'') \right) = 0 \; . \label{simple}
\end{equation}
However, appearances are deceiving: the unique solution to (\ref{simple}) is
$u(s') = 0$. To see this first note that $u(s') = 0$ does solve the equation. 
Now set $s' = 0$ and $s' = s$ in (\ref{simple}) to discover that $u(0) = 0 = 
u(s)$. This means that any solution must vanish at the endpoints. Suppose
that $u(s_1) = 0 = u(s_2)$ but that $u(s') \neq 0$ for $s_1 < s' < s_2$. By 
differentiating (\ref{simple}) with respect to $s'$,
\begin{equation}
u'(s') - \frac2{s} \int_0^s ds'' e A_-\left(x^+ - u(s'')\right) 
+ 2 e A_-\left(x^+ - u(s') \right) = 0 \; , \label{first}
\end{equation}
we see that $u'(s_1) = u'(s_2)$. If this common derivative is positive then 
$u(s')$ must rise from $s_1$ and subsequently fall back through zero so that 
it can rise at the same rate at $s_2$. If the common derivative is negative 
then the solution must fall from $s_1$ but subsequently rise back through 
zero so that it can fall at the same rate at $s_2$. Either way, there must 
be a zero for some $s_3$ between $s_1$ and $s_2$, which contradicts the 
assumption that $u(s') \neq 0$ between the two zeroes. Hence $u(s') = 0$ is 
the unique solution.

The Jacobian arising from the functional integral over $u(s')$ is the 
determinant of the following operator,
\begin{equation}
J(s',s'') = \delta(s' - s'') - 2 e E(x^+) \frac{s'}{s} + 2 e E(x^+) \theta(
s' - s'') \; .
\end{equation}
We can evaluate its determinant by first finding the inverse operator and then
using this to compute the derivative of the determinant with respect to $e E$. 
(To simplify the notation we drop the argument of the electric field.) To 
construct the inverse operator we must find the function $u(s')$ which obeys 
the integral equation,
\begin{eqnarray}
v(s') & = & \int_0^s ds'' J(s',s'') u(s'') \; , \\
& = & u(s') - 2 e E \frac{s'}{s}\int_0^s ds'' u(s'') + 2 e E \int_0^{s'} ds'' 
u(s'') \; .
\end{eqnarray}
Note first that the two functions agree at the endpoints: $u(0) = v(0)$ and 
$u(s) = v(s)$. Now differentiate twice with respect to $s'$,
\begin{equation}
v''(s') = u''(s') + 2 e E u'(s') \; .
\end{equation}
This is a first order differential equation for the derivative,
\begin{equation}
u'(s') = e^{-2 e E s'} u'(0) + e^{-2 e E s'} \int_0^{s'} ds'' e^{2 e E s''} 
v''(s'') \; .
\end{equation}
Integrating gives the function,
\begin{eqnarray}
\lefteqn{u(s') =} \nonumber \\
& & u(0) + \left({1 - e^{-2 e E s'} \over 2 e E}\right) u'(0) + \int_0^{s'} 
ds'' e^{-2 e E s''} \int_0^{s''} ds''' e^{2 e E s'''} v''(s''') \; , \quad \\
& & = v(s') + \left({1 - e^{-2 e E s'} \over 2 e E}\right) \left(u'(0) - 
v'(0) + 2 e E v(0) \right) \nonumber \\
& & \hspace{4cm} - 2 e E e^{-2 e E s'} \int_0^{s'} ds'' e^{2 e E s''} v(s'') 
\; .
\end{eqnarray}
We have already used the initial condition, $u(0) = v(0)$. Enforcing the final
condition, $u(s) = v(s)$, implies,
\begin{equation}
u'(0) - v'(0) + 2 e E v(0) = (2 e E)^2 \left({e^{-e E s} \over e^{e E s} - 
e^{-e E s}}\right) \int_0^s ds'' e^{-2 e E s''} v(s'') \; .
\end{equation}
The complete solution is therefore,
\begin{eqnarray}
\lefteqn{u(s') = v(s') - 2 e E e^{-2 e E s'} \int_0^{s'} ds'' e^{2 e E s''} 
v(s'')} \nonumber \\
& & \hspace{3cm} + 2 e E e^{-e E (s + s')} {\sinh(e E s') \over \sinh(e E s)} 
\int_0^s ds'' e^{2 e E s''} v(s'') \; . \label{ufromv}
\end{eqnarray}

Functionally differentiating (\ref{ufromv}) with respect to $v(s'')$ gives
the inverse of the Jacobian operator,
\begin{eqnarray}
\lefteqn{J^{-1}(s',s'') = \delta(s' - s'') - 2 e E e^{-2 e E (s'-s'')} 
\theta(s'-s'')} \nonumber \\
& & \hspace{2cm} + 2 e E e^{-e E (s + s' - 2 s'')} {\sinh(e E s') \over 
\sinh(e E s)} \; .
\end{eqnarray}
Now differentiate the logarithm of the determinant of $J$ with respect to $eE$,
\begin{eqnarray}
{\partial \over \partial eE} \ln\left({\rm det}[J]\right) & = & \int_0^s ds'
\int_0^s ds'' \left[-2 \frac{s'}{s} + 2 \theta(s'-s'')\right] J^{-1}(s'',s')
\; , \\
& = & -\frac1{eE} + s \coth(eEs) \; .
\end{eqnarray}
Integrating and making use of the fact that ${\rm det}[J] = 1$ for $eE = 0$
gives,
\begin{equation}
{\rm det}[J] = {\sinh\left(e E(x^+) s\right) \over e E(x^+) s} \; .
\end{equation}

It is now simple to write down the one loop effective action,
\begin{eqnarray}
\Gamma_1[A] & = & -i L[A] \; , \\
& = & {1 \over 8 \pi^2} \int d^4x \int_0^{\infty} {ds \over s^3} e^{-i s m^2} 
\left\{e E(x^+) s \coth\left(e E(x^+)s\right) - 1 \right\} \; .
\end{eqnarray}
This has exactly the same form as Schwinger's result \cite{Schwinger} except
that our electric field can depend arbitrarily upon the light cone time
parameter $x^+$. The real part has the same (universal) one loop divergence,
whereas the imaginary part can be evaluated by first extending the range of
integration and then closing the contour above and below,
\begin{eqnarray}
\lefteqn{2 {\rm Im}\left(\Gamma[A] \right)} \nonumber \\
& & = -{1 \over 8 \pi^2} \int d^4x \int_{-\infty}^{\infty} {ds \over s^3} 
\sin(s m^2) \left\{e E(x^+) s \coth\left(e E(x^+)s\right) - 1 \right\} \; , \\
& & = \int d^4x \left\{- {e^2 E^2(x^+) \over 24 \pi} + {1 \over 4 \pi} 
\sum_{n=1}^{\infty} \left({ e E(x^+) \over n \pi}\right)^2 \exp\left[- {n \pi 
m^2 \over \vert e E(x^+) \vert}\right] \right\} \; . \quad \label{P1}
\end{eqnarray}

The second term in the brackets of (\ref{P1}) agrees (for constant E) with
Schwinger's result for the volume rate of pair production \cite{Schwinger}. 
He avoided the first term by making an additional subtraction to remove the 
ultraviolet divergence,
\begin{equation}
\Gamma_R[A] \equiv {1 \over 8 \pi^2} \int d^4x \int_0^{\infty} {ds \over s^3} 
e^{-i s m^2} \left\{e E s \coth\left(e E s\right) - 1 - {(e E s)^2 \over 3}
\right\} \; .
\end{equation}
We can see from (\ref{P1}) that the subtracted term contributes a nonzero
imaginary part which cannot properly belong to a counterterm. That Schwinger 
was nevertheless correct to ignore this term in the volume rate of pair 
production is proven by its absence in the real time, operator computation 
\cite{TTW1}.

The fact that our effective action happens to have the same form as Schwinger's
should not detract from the enormous generalization it represents. It is also
significant as an exact calculation which contains an essential singularity at 
$e=0$. Other, recent calculations of analogous pair production processes also 
show essential singularities \cite{HoNa,FGMA}, as do instanton approximations 
of the vacuum structure in a variety of theories \cite{MeOo,LM}. These 
essential singularities seem to confirm Dyson's famous observation \cite{Dyson} 
that QED cannot be analytic at $\alpha = 0$.

\vskip .5cm
\centerline{\bf Acknowledgements}

This work was partially supported by DOE contract DE-FG02-97ER\-41029 and by
the Institute for Fundamental Theory at the University of Florida.


\begin{thebibliography}{99}

\bibitem{Schwinger} J. Schwinger, Phys. Rev. {\bf 82}, 664 (1951).

\bibitem{HoNa} M. N. Hounkonnou and M. Naciri, J. Phys. {\bf G26}, 1849 (2000).

\bibitem{FGMA} H. M. Fried, Y. Gabellini, B. H. J. McKellar and J. Avan, Phys.
Rev. {\bf D63}, 125001 (2001).

\bibitem{TTW1} T. N. Tomaras, N. C. Tsamis and R. P. Woodard, Phys. Rev. {\bf 
D62}, 125005 (2000), hep-th/0007166.

\bibitem{TTW2} T. N. Tomaras, N. C. Tsamis and R. P. Woodard, ``Pair creation 
and axial anomaly in light-cone $QED_2$,'' hep-th/0108090.

\bibitem{Fried1} H. M. Fried, {\it Functional Methods and Models in Quantum 
Field Theory}, (MIT Press, Cambridge, 1972).

\bibitem{Fradkin} E. S. Fradkin, Nucl. Phys. {\bf 76}, 588 (1966).

\bibitem{Fried2} H. M. Fried, {\it Functional Methods and Eikonal Models},
(Editions Frontieres, Gif-sur-Yvette, 1990).

\bibitem{MeOo} P. F. Mende and H. Ooguri, Nucl. Phys. {\bf B39}, 641 (1990).

\bibitem{LM} J. Lee and P. F. Mende, Phys. Lett. {\bf B312}, 433 (1993), 
hep-th/9211049.

\bibitem{Dyson} F. J. Dyson, Phys. Rev. {\bf 85}, 631 (1952).

\end{thebibliography}
\end{document}